\RequirePackage{fixltx2e} 
\documentclass[10pt,superscriptaddress,aps,twocolumn,prl,nobalancelastpage,raggedbottom]{revtex4-1}

\usepackage{graphicx}
\usepackage[usenames,dvipsnames]{color}
\usepackage{siunitx} 
\usepackage[a4paper]{geometry} 
\usepackage{amsmath}                    
\usepackage{amssymb}                    
\usepackage{subfigure}                  
\usepackage[version=3]{mhchem}          
\usepackage[T1]{fontenc}                
\usepackage{textcomp}

\begin{document}

\title{Shaping, imaging and controlling plasmonic interference fields at buried interfaces}

\author{Tom T. A. Lummen}\email{tom.lummen@epfl.ch}
\affiliation{Laboratory for Ultrafast Microscopy and Electron Scattering, ICMP, \'{E}cole Polytechnique F\'{e}d\'{e}rale de Lausanne, Station 6, CH-1015 Lausanne, Switzerland}
\author{Raymond J. Lamb}
\affiliation{SUPA, School of Physics and Astronomy, University of Glasgow, Glasgow G12 8QQ, United Kingdom}
\author{Gabriele Berruto}
\affiliation{Laboratory for Ultrafast Microscopy and Electron Scattering, ICMP, \'{E}cole Polytechnique F\'{e}d\'{e}rale de Lausanne, Station 6, CH-1015 Lausanne, Switzerland}
\author{Thomas LaGrange}
\affiliation{Interdisciplinary Center for Electron Microscopy (CIME), \'{E}cole Polytechnique F\'{e}d\'{e}rale de Lausanne, CH-1015 Lausanne, Switzerland}
\author{Luca Dal Negro}
\affiliation{Department of Electrical and Computer Engineering and Photonics Center, Boston University, 8 Saint Mary\textquotesingle s Street, Boston, Massachusetts 02215, United States}
\author{F. Javier Garc\'{i}a de Abajo}
\affiliation{ICFO - Institut de Ciencies Fotoniques, The Barcelona Institute of Science and Technology, 08860 Castelldefels, Barcelona, Spain}
\altaffiliation{ICREA-Instituci\'{o} Catalana de Recerca i Estudis Avancats, Passeig Llu\'{i}s Companys, 23, 08010 Barcelona, Spain}
\author{Damien McGrouther}
\affiliation{SUPA, School of Physics and Astronomy, University of Glasgow, Glasgow G12 8QQ, United Kingdom}
\author{Brett Barwick}
\affiliation{Department of Physics, Trinity College, 300 Summit St., Hartford, Connecticut 06106, United States}
\author{Fabrizio Carbone}\email{fabrizio.carbone@epfl.ch}
\affiliation{Laboratory for Ultrafast Microscopy and Electron Scattering, ICMP, \'{E}cole Polytechnique F\'{e}d\'{e}rale de Lausanne, Station 6, CH-1015 Lausanne, Switzerland}

\begin{abstract}
Filming and controlling plasmons at buried interfaces with nanometer (nm) and femtosecond (fs) resolution has yet to be achieved and is critical for next generation plasmonic/electronic devices.
In this work, we use light to excite and shape a plasmonic interference pattern at a buried metal-dielectric interface in a nanostructured thin film. Plasmons are launched from a photoexcited array of nanocavities and their propagation is filmed via photon-induced near-field electron microscopy (PINEM). The resulting movie directly captures the plasmon dynamics, allowing quantification of their group velocity at $\simeq$ 0.3c, consistent with our theoretical predictions. Furthermore, we show that the light polarization and nanocavity design can be tailored to shape transient plasmonic gratings at the nanoscale. These results, demonstrating dynamical imaging with PINEM, pave the way for the fs/nm visualization and control of plasmonic fields in advanced heterostructures based on novel 2D materials such as graphene, \ce{MoS2}, and ultrathin metal films. 
\end{abstract}

\date{\today}

\maketitle

\newpage


\section{Introduction}
Surface plasmon polaritons (SPPs)\cite{maier07} are considered ideal for circuitry miniaturization to nanoscale dimensions\cite{barnes03}, due to their inherent surface-bound and spatial confinement characteristics. This makes them appealing for various applications, such as nanoscopic signal processing\cite{ozbay06} and high resolution sensing\cite{stewart08}. However, considering the fs temporal and nm spatial scales involved, direct real space characterization of evanescent waves presents a significant experimental challenge, especially in materials and devices involving buried interfaces and complex heterostructures.\newline
\indent Several optics-based approaches have been developed to meet the challenge of tracking surface-bound SPPs with appropriate resolution, notably dual-color coherent anti-Stokes Raman scattering (CARS)\cite{liu12} and scanning tip based techniques such as near-field scanning optical microscopy (NSOM) \cite{bohm94,gersen05}. In recent years, the spatial and temporal resolution limits of these optical techniques have been pushed to the 10s of nm and 100s of fs \cite{wagner14,nishiyama15_2}. In an alternative approach, the use of electron photoemission for detection, as in time-resolved nonlinear photoemission electron microscopy (TR PEEM)\cite{petek97,kubo05,kubo07}, has allowed for comparable spatial resolution in imaging SPPs, with time resolution extending down to the few-fs single optical cycle regime. Recent advances in TR-PEEM methods have facilitated the successful spatiotemporal mapping of localized and propagating surface plasmons in a variety of systems\cite{aeschlimann10,lemke13,lemke14,gong15,marsell15}. However, as TR-PEEM relies on sample-emitted photoelectrons, and scanning-tip-based techniques require physical probe proximity to the evanescent field, both are inherently limited to the study of exposed material surfaces. Furthermore, the connection between the probed field and the local density of photonics states is not always direct\cite{bauer07,lemke12}. Due to these restrictions, these techniques are not suited for studies of advanced multilayer systems and heterostructured devices that rely on plasmonic waves bound to buried interfaces.\newline
\indent An alternative method to those mentioned above employs electron energy loss spectroscopy (EELS) in a transmission electron microscope\cite{Kociak2011}, either in a spectrum imaging (SPIM) or energy-filtered transmission electron microscopy (EFTEM) configuration, enabling $\sim$ 1 nm spatial resolution in the imaging of SPPs in thin films and nanostructures\cite{bosman07,nelayah07,schaffer09}. In these experiments, plasmonic guided modes are excited by the transient electric field generated by the fast (80-300 keV) electrons traversing the specimen\cite{garcia04,rossouw13}. In its most widespread SPIM implementation, EELS is combined with an electron beam raster scanning approach, offering temporal imaging resolution down to the millisecond (camera-rate) range. Recently however, a new technique termed photon-induced near-field electron microscopy (PINEM)\cite{barwick09} extended the accessible domain of EELS-based experiments to the nm/fs regime by optically exciting SPPs with intense fs laser pulses, and subsequently probing the resulting plasmonic near-fields through their interaction with ultrafast electron bunches\cite{garcia10_2,park10,park14}. By creating an image using only the electrons that have gained energy through the inelastic scattering from plasmons, PINEM performed in the EFTEM configuration enables direct field-of-view imaging of photoexcited plasmonic fields in a transmission geometry\cite{piazza13,barwick15}. Compared to CARS, NSOM, and TR-PEEM, the PINEM technique offers several major advantages: \textit{i)} it allows experimental access to otherwise "hidden" plasmonic fields at interfaces buried beneath the surface of advanced multilayer designs; \textit{ii)} it uses a noninvasive probe that does not strongly affect the plasmons being measured; and \textit{iii)} it couples exclusively to the evanescent component of the electromagnetic field at the sample, avoiding incident or reflected beam signal contributions and enabling a direct image interpretation. These advantages, along with the convenient ability to obtain complementary nanoscale morphological, structural and chemical information through  static or time-resolved TEM imaging, electron diffraction and EELS in the same setup, make PINEM uniquely suited for probing plasmonics in more complex geometries and advanced multilayer systems. In recent years, PINEM has been used to image optical near-fields at the surfaces of a variety of materials, nanostructures and biological specimens\cite{barwick09,piazza13,piazza15,barwick15,feist15}, although these efforts did not track the dynamical behavior of the photoinduced plasmons or target the material properties of the specimens. So far, the use of PINEM to experimentally access SPP dynamics has remained an open challenge due to the typical SPP lifetime (on the order of 100 fs) being beyond instrumental temporal resolution (to date at best $\sim$ 500 fs using single optical pulses\cite{piazza13}, extendable down to $\sim$ 210 fs through photon gating\cite{Hassan2015}), with the latter being primarily limited by the electron bunch duration\cite{barwick15}.\newline
\indent In this work we employ localized, embedded SPP sources to launch SPPs in a 2D plasmonic waveguide, provided by a buried Ag/\ce{Si3N4} interface, allowing the photoinduced plasmons to propagate and interfere over several microns. In this geometry, the electron-plasmon scattering observed at different distances from the SPP sources occurs at different times due to the finite time delay needed for SPP propagation. By visualizing and characterizing this temporal dependence using time-resolved PINEM imaging, we could film the dynamical evolution of the collective plasmonic field propagating at a buried interface with combined ultrafast and nanoscale resolution.\newline
\onecolumngrid
\begin{center}
\begin{figure}[htb]
\includegraphics[width=12cm]{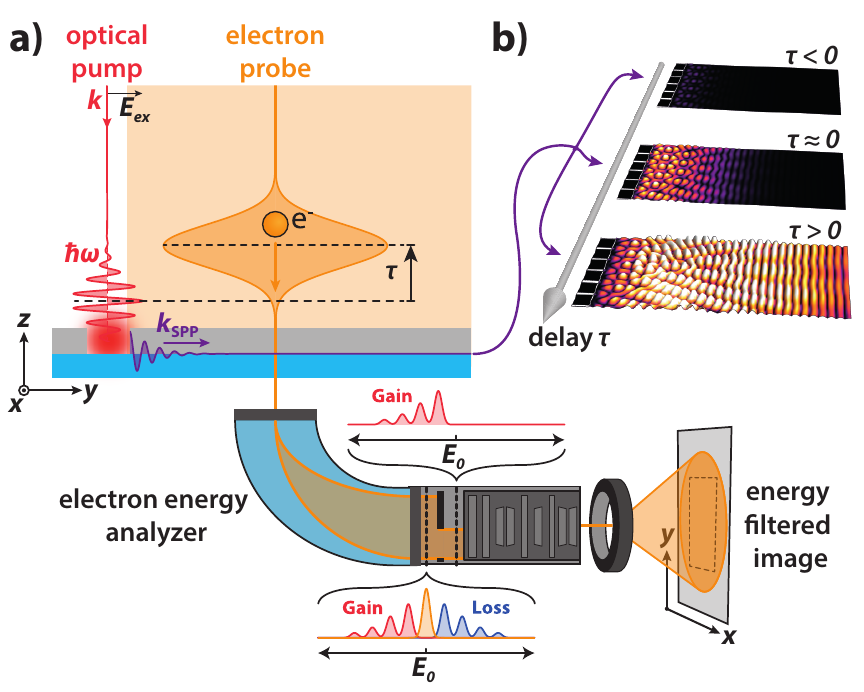}
\caption{\label{Fig1} \textbf{Time-resolved PINEM methodology.} \textbf{(a)} Simplified scheme of the time-resolved PINEM experiments in this work. A photon pump pulse incident on a nanopatterned Ag-on-\ce{Si3N4} bilayer structure generates an SPP wave propagating along the buried Ag/\ce{Si3N4} interface. The near-field of the propagating SPP is subsequently probed through its interaction with a field-of-view electron pulse at a time-delay $\tau$. Energy-filtered imaging of the resulting electron distribution of transmitted electrons then provides spatially-resolved temporal snapshots of the near-field corresponding to the propagating plasmonic wave. \textbf{(b)} Variation of the relative time delay between the optical excitation pulse and the probing electron pulse generates a time-resolved movie of the ultrafast evolution of the buried plasmonic near-field.}
\end{figure}
\end{center}
\twocolumngrid
\section{SPP propagation}
\indent A conceptual scheme of our experiment is sketched in Fig. 1a. A linearly polarized optical pump pulse is incident on a nanocavity in a \ce{Si3N4}-supported Ag thin film, launching SPPs of energy $E_{\text{SPP}}$ = $\hbar\omega$ that are confined to the Ag/\ce{Si3N4} interface. At a variable time delay $\tau$, the propagating plasmonic wave is then probed by a spatially dispersed pulse of fast electrons, which inelastically scatter from the optically excited SPPs. This scattering interaction results in the electrons gaining and losing integer numbers of SPP energy quanta\cite{garcia10_2,park10,asenjo13,feist15}, yielding equidistantly spaced peaks in the energy distribution of the transmitted electrons. By subsequently constructing an energy-filtered image using only those electrons that have gained energy, one directly obtains a spatially-resolved snapshot of the $E_{z}$ component of the buried plasmonic near-field\cite{barwick09,barwick15}. Due to the large field-of-view of the electron beam employed here, snapshots recorded at different relative time delays $\tau$ thus capture the spatiotemporal evolution of the photoinduced plasmonic wave (see Fig. 1b).\newline
\indent The sample studied in this work consists of a 30 nm Ag thin film deposited on a 50 nm \ce{Si3N4} membrane. A variety of different rectangular nanocavity (NC) arrays were patterned into the Ag film using focused ion beam milling (see Methods and Supplementary Figure 1). In order to experimentally measure the propagation speed of SPPs at the buried interface, we monitor a large, featureless area adjacent to a linear array of NCs, as sketched in Fig. 2a. The optical pump beam ($\varnothing$ $\simeq$ \SI{100}{\micro\metre}, $\lambda_{\text{0}}$ = 786 nm, full-width-half-maximum (FWHM) duration $\sim$ 105 fs) and the photoelectron probe beam ($\varnothing$ $\simeq$ 40 \SI{}{\micro\metre}), $E_{0}$ = 200 keV, FWHM duration $\sim$ 650 fs) are spatially overlapped at near-normal incidence on the Ag side of the nanopatterned bilayer sample. The optical pump pulses generate SPPs at the edges of the NCs, with each edge coupling to the incident electric field component along the in-plane edge normal\cite{lalanne06}. Since the NCs fully perforate the metal film, photoexcited SPPs are transmitted through the NCs to the other side of the Ag layer and subsequently launched across the Ag/\ce{Si3N4} interface\cite{lin13}. In first approximation, each NC edge thus effectively acts as a source of SPPs that travel radially outward in a point dipole-like pattern at the buried interface\cite{tanemura11}. Consequently, according to the Huygens-Fresnel principle, the interferometric plasmonic wave propagating away from the NC array (along the y-axis in Fig. 2a) corresponds to the coherent superposition of the SPPs launched from the different NCs.\newline
\indent Figure 2b shows the experimental PINEM image of the resulting plasmonic interference pattern (PIP), recorded at a zero relative delay time $\tau$ between the optical pump and electron probe pulse maxima. The spatial fast Fourier transform (FFT) of the entire frame in Fig. 2b gives direct access to the spatial frequencies present in the interference pattern\cite{gersen05}, and its radial integral is singularly peaked at the spatial period corresponding to the wavelength of the optically driven SPP wave (Fig. 2c). We can thus directly extract the SPP wavelength $\lambda_{\text{SPP, exp}}$ and wavevector $k_{\text{SPP}} = 2\pi/\lambda_{\text{SPP, exp}}$ as 633 $\pm$ 13 nm and 9.94 $\pm$ 0.20 $\mu$m$^{-1}$, respectively, where the errors are due to the uncertainty in the calibration of the PINEM image (see Methods). Figure 2d depicts the calculated electron energy loss probability for 200 keV electrons transmitted through a model of our layered sample (30 nm Ag on 50 nm \ce{Si3N4}), plotted as a function of energy loss and momentum transfer normal to the beam direction. Given the known energy of the optically driven SPPs in the measurements ($E_{\text{SPP}}$ = $E_{\text{photon}}$ = 1.58 $\pm$ 0.015 eV), we find that the experimentally obtained SPP characteristics match well with the theoretical dispersion of the plasmonic mode propagating at the buried interface (see Supplementary Figure 2).\newline
\indent Since PINEM is an ultrafast field-of-view technique, it can track the evolution of the photoexcited plasmonic wave in time, given a sufficient combined spatiotemporal resolution. Here we can follow its evolution and build-up as it propagates away from the linear NC array upon photoexcitation by performing a slice analysis of the PINEM images obtained at different time delays $\tau$. As sketched in Figure 3a, we divide up the image in a series of horizontal slices centered at different distances $y$ from the NC array. By taking the spatial FFT of these slices for each delay, and integrating the Fourier features corresponding to the SPP interference, we can track the propagation of the plasmonic wavefront in time, as is sketched for a number of discrete slices in Fig. 3b. These time traces correspond to the sub-ps cross-correlation of the electron probe and the photoexcited SPP pulse in the different discrete $y$-ranges in the sample, offset by a temporal difference $\Delta t$ due to the SPP wave propagation. Combining the data of all slices yields a quantitative description of the dynamic behavior, which directly captures the SPP propagation. Figure 3c depicts the temporal center of Gaussian fits to the cross-correlation time traces of different slicing schemes as a function of the distance $y$ from the NC array. The linear slope of this curve is a direct measure of the group velocity of the composite plasmonic wave, corresponding to approximately a third of the free-space speed of light, $v_{\text{g, exp}}$ = (9.4 $\pm$ 1.3) $\times$ 10$^7$ m s$^{-1}$. This value is consistent with that expected from theory for the buried SPP mode, as indicated in Figure 2d. Although SPP group velocities at surfaces have been previously extracted from both time domain\cite{bai04} and frequency domain\cite{temnov07} experiments, this experiment demonstrates the direct measurement of the SPP propagation speed at a buried interface.\newline
\onecolumngrid
\begin{center}
\begin{figure}[h]
\includegraphics[width=11cm]{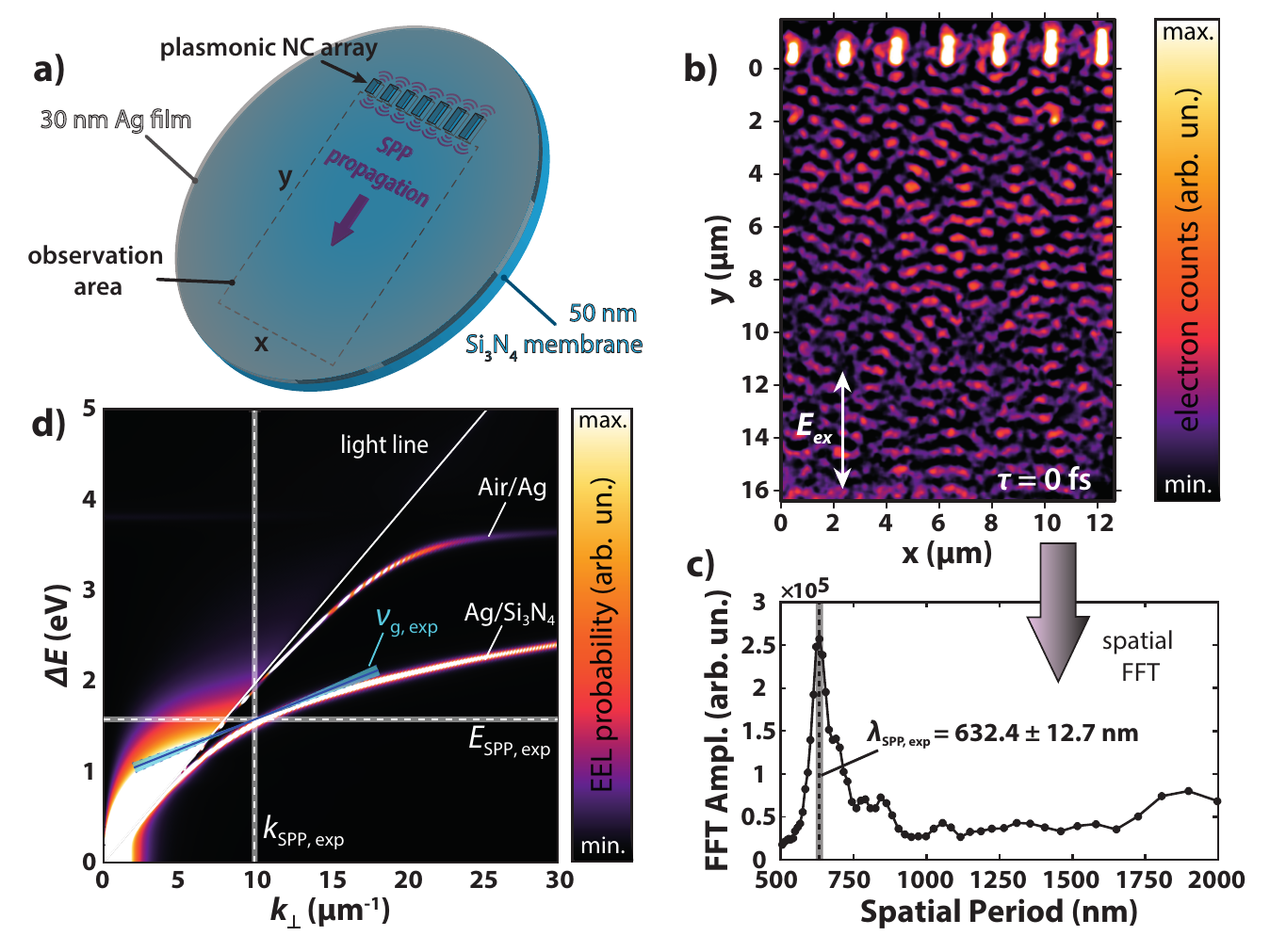}
\caption{\label{Fig2} \textbf{PINEM imaging of SPP interference at a buried interface.} \textbf{(a)} Schematic of the sample and SPP propagation experiment. \textbf{(b)} Experimental PINEM image of the photoinduced plasmonic wave propagating at the buried Ag/\ce{Si3N4} interface. The image was recorded at zero delay between electron and laser pulses ($\tau$ = 0 ps), using only electrons that have gained energy from the plasmonic near-field. The excitation light was polarized parallel to the long axis of the NCs. As the electron and optical pulse durations satisfy $\Delta_{e}$ $\gg$ $\Delta_{p}$, the plasmonic interference pattern (PIP) is observed in the entire window at this delay. The linear false color scale corresponds to relative electron counts. \textbf{(c)} Radially integrated fast Fourier transform (FFT) amplitude of the spatial frequency components contained in the PIP of panel b as a function of the corresponding spatial period (i.e. the radial distance in Fourier space). The position of the singular peak, indicated by the dashed line, corresponds to $\lambda_{\text{SPP, exp}}$, and its uncertainty is indicated by the shaded gray line. \textbf{(d)} Analytically calculated\cite{garcia04} electron energy loss (EEL) probability of 200 keV electrons traversing a 30 nm Ag on 50 nm \ce{Si3N4} layer stack (see panel a) at normal incidence, as a function of energy loss $\Delta E$ and transversal momentum transfer $k_{\perp}$ (with respect to the beam direction). Two dispersion branches are observable to the right of the light cone (solid straight line), corresponding to SPP modes propagating along the air/Ag and Ag/\ce{Si3N4} interfaces, respectively (see Supplementary Figure 2). Experimental magnitudes are indicated by dashed lines (see text), with corresponding uncertainties indicated by thick shaded lines. In particular, the solid blue line indicates the measured SPP group velocity $v_{\text{g, exp}}$ (see Fig. 3), with the shaded hourglass shape representing the corresponding experimental error.}
\end{figure}
\end{center}
\twocolumngrid
\onecolumngrid
\begin{center}
\begin{figure}[t]
\includegraphics[width=13cm]{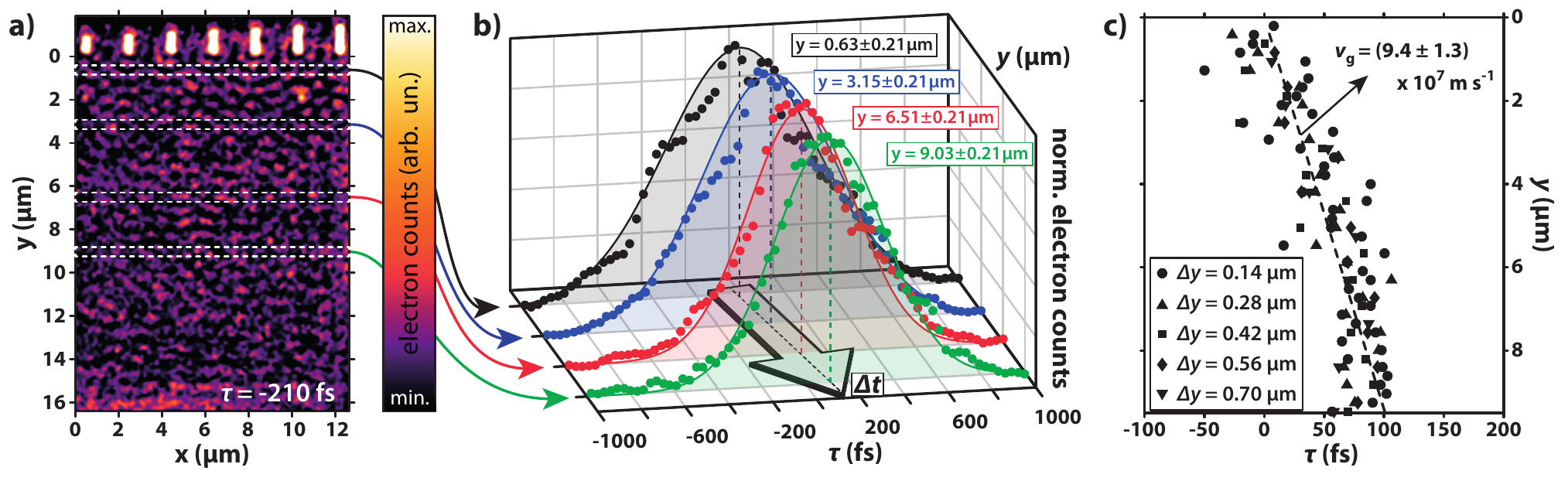}
\caption{\label{Fig3} \textbf{Direct PINEM measurement of SPP propagation.} \textbf{(a)} Experimental PINEM image of the photoinduced plasmonic wave propagating at the buried Ag/\ce{Si3N4} interface. The image was recorded at $\tau$ = -210 fs, using only electrons that have gained energy from the plasmonic near-field. White dashed rectangles exemplify the discrete slicing of the image for spatial Fourier analysis. Electron counts are plotted on a linear false color scale. \textbf{(b)} Temporal traces of the SPP wave intensity in the different discrete slices indicated in a. For each slice at each time delay $\tau$, the SPP wave intensity is calculated as the integrated intensity of the corresponding features in the spatial FFT. The $y$-range of each of the slices corresponds to $\Delta y$ = \SI{0.42}{\micro\metre}. \textbf{(c)} Temporal center positions of Gaussian fits to the cross-correlation time traces versus distance traveled by the SPP wave ($y$). Data points corresponding to different slicing schemes (varying $\Delta y$) are combined (see Methods). The linear slope fitted to the combined data (dashed line) directly yields the group velocity of the SPP wave, $v_{\text{g, exp}}$ = (9.4 $\pm$ 1.3) $\times$ 10$^7$ m s$^{-1}$, corresponding to an effective SPP refractive index of c/$v_{\text{g, exp}}$ $\simeq$ 3.}
\end{figure}
\end{center}
\twocolumngrid
\section{Shaping transient plasmonic interference patterns}
\indent The ability of PINEM to directly access plasmonic interference patterns opens up a multitude of possibilities for creating arbitrarily shaped plasmonic transient gratings, as the obtained interferometric structures can be precisely tailored by combining appropriately arranged nanoscale features with suitable photoexcitation\cite{tanemura11,lin13}. Moreover, experiments can be supported by simulations applying Maxwell's equations to the nanopattern architectures, providing effective feedback and enabling the predictive design of transient PIPs. Figure 4 illustrates this flexibility in designing transient PIPs. Panels a and b show the PINEM image of a linear array of long NCs under photoexcitation using different linear polarization orientations, demonstrating that it is indeed possible to obtain transient plasmonic gratings using relatively simple nanostructures. More complex interference patterns can be generated away from such linear NC arrays, as shown for example for an array formed by short NCs (Fig. 4c). One particularly interesting feature of an SPP wave propagating away from a periodic source array is the plasmon Talbot effect, in which the source configuration is self-imaged into an array of plasmon focal spots at characteristic propagation distances\cite{dennis07}. This is predicted to occur even outside the paraxial regime, and upon close inspection its signature can indeed be seen in Figure 2b at $y \approx$ \SI{4.2}{\micro\meter}, corresponding to the half-period shifted revival at half the so-called Talbot distance. By combining two linear NC arrays with an appropriate relative positioning and photoexcitation one can exploit this effect even at shorter propagation distances, allowing for the generation of multiple closely spaced arrays of nanoscale focal spots of near-field intensity, and thus charge density, at the buried \ce{Ag/Si3N4} interface (see Fig. 4d). Here the PIP is dominated by the interference of counter-propagating SPPs generated at the different arrays, such that its $\lambda_{\text{SPP}}$/2 periodicity along the propagation axis yields $\lambda_{\text{SPP, exp}}$ = 638 $\pm$ 32 nm, which is consistent with the value extracted in Figure 2. Given the thin metal layer and relatively narrow NCs, the influence of multiple scattering and reflections of the SPPs on the structure of the resulting standing wave pattern is small, such that a simple analytical model based on a double array of point dipole SPP sources can reproduce the major features observed (see Supplementary Figure 3). By contrast, this model does not capture the observed complex interference of co-propagating SPPs (as in the propagation experiment). In this case, the time-averaged interference between SPPs from the different point-dipole sources is weaker (similar to that in the plasmon Talbot effect), and additional experimental factors such as non-uniform time-averaging (due to the optical pulse envelope), finite bandwidth excitation and multiple order PINEM scattering are expected to play a dominant role.\newline
\indent As shown in Fig. 4a-b, the shape of the transient SPP grating generated by the linear array of long NCs is strongly dependent on the excitation polarization. Figure 5 analyzes the indicated region in more detail. In particular, we show in Fig. 5a that upon photoexcitation using light polarized perpendicular to the NC long axis, the counterpropagating SPPs between the NCs set up a transient plasmonic grating with a period corresponding to half the SPP wavelength. This yields another independent determination of the SPP carrier wavelength ($\lambda_{\text{SPP, exp}}$ = 639 $\pm$ 6 nm), in good agreement with both previously extracted values and the calculated dispersion for the SPP mode propagating at the \ce{Ag/Si3N4} interface. Interestingly, varying the polarization angle of the incident light dramatically changes the relative intensities of the near-field grating fringes. At a polarization angle of $\varphi=20^{\circ}$, two of the grating fringes are almost completely suppressed, such that the periodicity of the plasmonic grating has effectively doubled. This is further illustrated in Figs. 5b and c, which depict the full polarization dependence of the transient plasmonic grating. Panel b plots the fringe contrast $\eta$, calculated as $I_{\text{PINEM, max}}$/$I_{\text{PINEM}}(\varphi)$, for each of the four fringes indicated in Fig. 5a, revealing a clear intensity dip in the $\varphi$-dependence of peaks 1 and 3 centered around $\varphi=20^{\circ}$. Figure 5c depicts a Fourier analysis of the SPP grating, with the dominant spatial period of the grating indicated by the red squares (right vertical axis) and the grating amplitude ratio $\rho$ (defined as the ratio of FTT amplitudes A(p$_1$)/A(p$_2$) at spatial periods p$_i$ = $\lambda_{\text{SPP}}/i$) indicated by the solid black circles (left vertical axis). These results demonstrate that it is possible to reversibly switch the periodicity of the transient plasmonic grating by tuning the light polarization, showing great promise as a potential means for ultrafast switching of localized charge density in quasi-two-dimensional (quasi-2D) materials and heterostructures.
\onecolumngrid
\begin{center}
\begin{figure}[htbp]
\includegraphics[width=10cm]{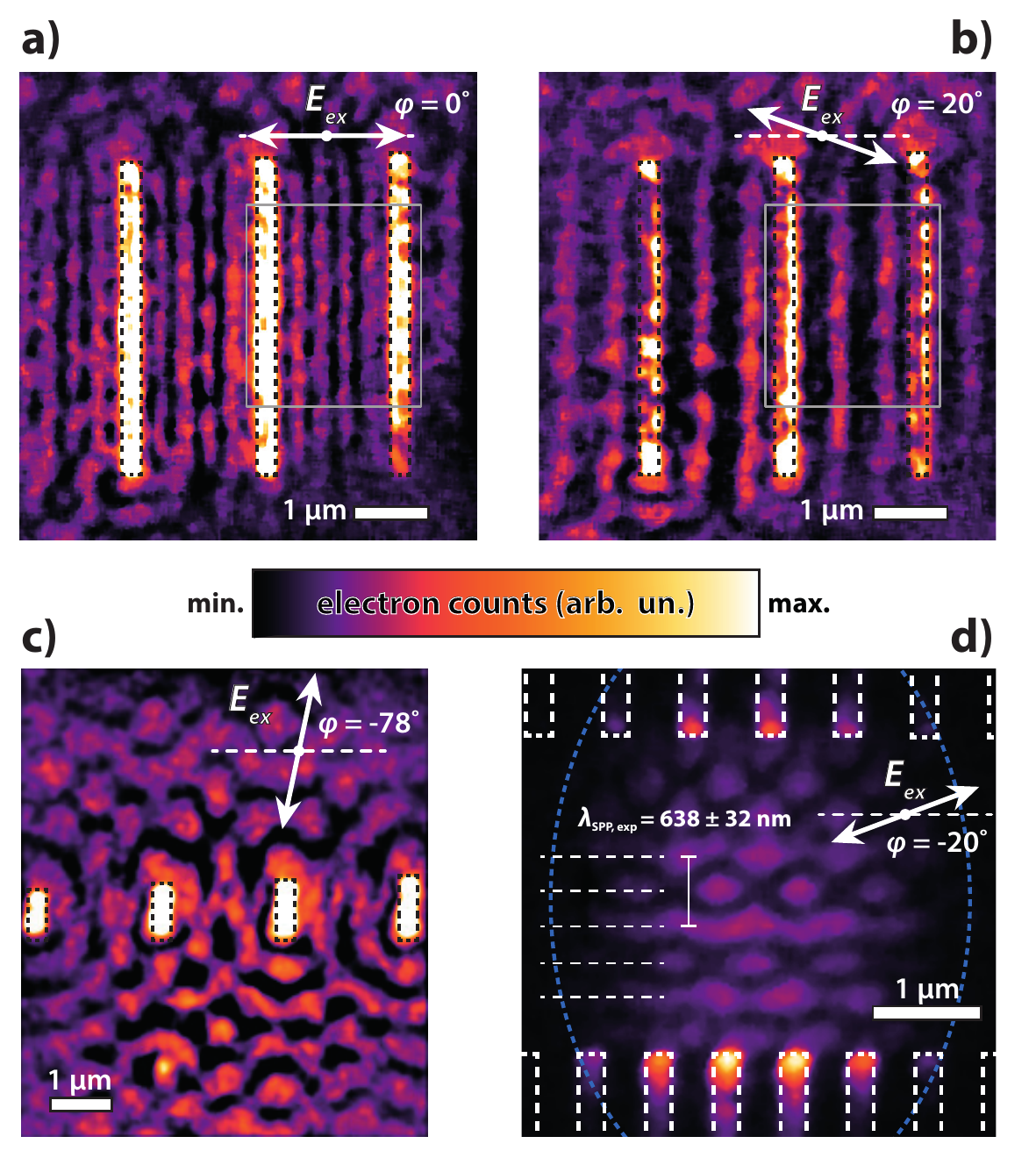}
\caption{\label{Fig4} \textbf{PINEM imaging of NC-based nanopatterns.} PINEM images of the transient SPP grating generated by a linear array of NCs (measured lengths $\simeq$ 4.2, 4.3 and \SI{4.4}{\micro\meter}, respectively, widths $\simeq$ 260 nm), photoexcited by linearly polarized light at $\varphi=0^{\circ}$ and $\varphi=20^{\circ}$, are shown in \textbf{(a)} and \textbf{(b)}, respectively. The area enclosed by the gray rectangle is analyzed in detail in Fig. 5. \textbf{(c)} PINEM image of the PIP generated by a linear array of short NCs (measured lengths $\simeq$ 0.6, 0.7, 0.8 and \SI{0.9}{\micro\meter}, respectively, widths $\simeq$ 275 nm), photoexcited by linearly polarized light at $\varphi=-78^{\circ}$. \textbf{(d)} PINEM image of the PIP generated by two vertically offset linear arrays of NCs (length $\simeq$ \SI{2.1}{\micro\meter}, width $\simeq$ 270 nm), photoexcited by linearly polarized light at $\varphi=-20^{\circ}$. The resulting SPP standing wave interference shows a periodicity of $\lambda_{\text{SPP}}$/2 along the vertical (counter-)propagation direction (dashed horizontal lines), yielding an estimated SPP wavelength of $\lambda_{\text{SPP, exp}}$ = 638 $\pm$ 32 nm. The dashed elliptical shape depicts the approximate outline of the electron beam footprint. All panels are recorded at $\tau =$ 0 using $\lambda_{\text{0}}$ = 786 nm and share the linear false color scale depicted in the middle.}
\end{figure}
\end{center}
\twocolumngrid
\onecolumngrid
\begin{center}
\begin{figure}[htbp]
\includegraphics[width=8cm]{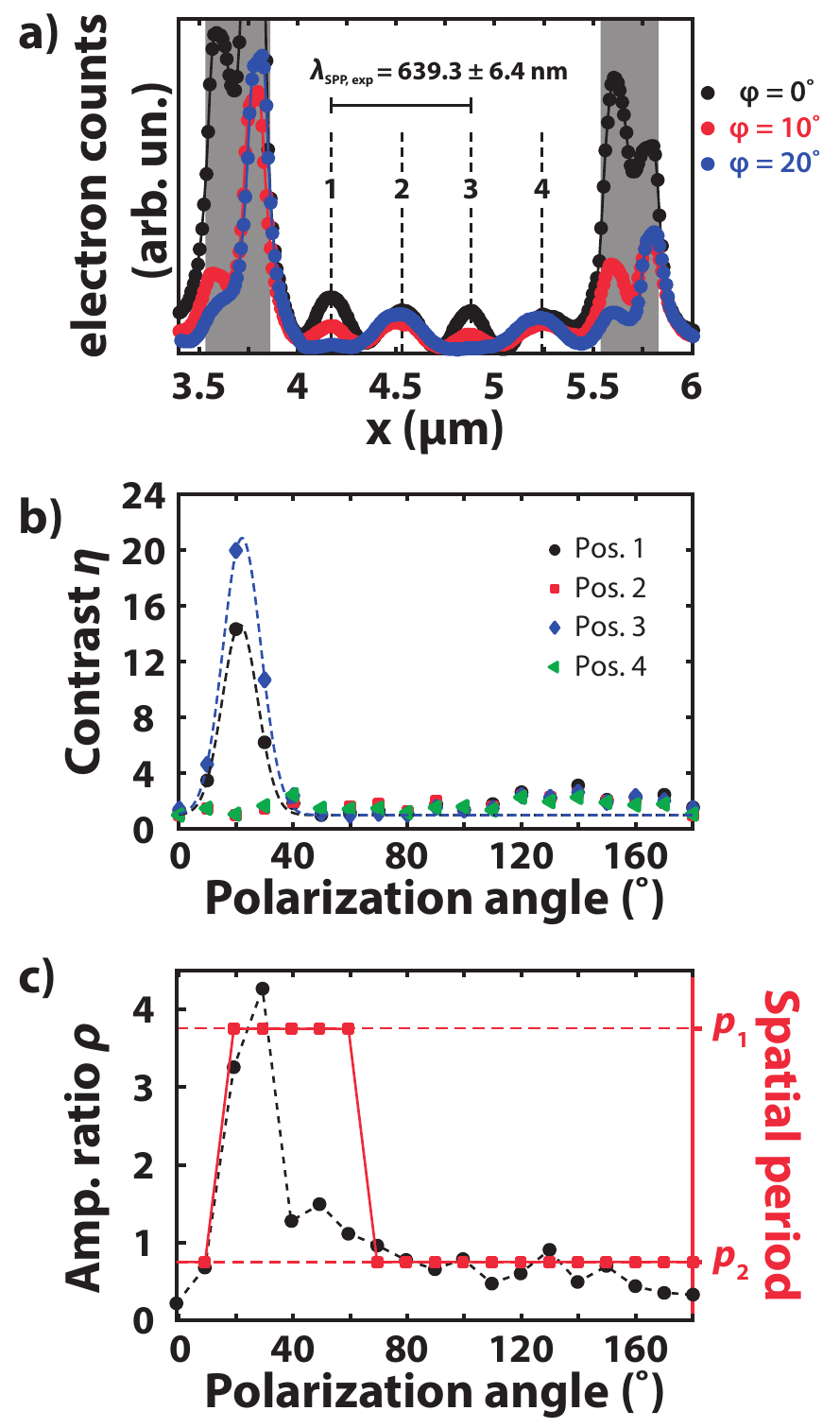}
\caption{\label{Fig5} \textbf{PINEM imaging of NC-based nanopatterns.} PINEM images of the transient SPP grating generated by a linear array of NCs (measured lengths $\simeq$ 4.2, 4.3 and \SI{4.4}{\micro\meter}, respectively, widths $\simeq$ 260 nm), photoexcited by linearly polarized light at $\varphi=0^{\circ}$ and $\varphi=20^{\circ}$, are shown in \textbf{(a)} and \textbf{(b)}, respectively. The area enclosed by the gray rectangle is analyzed in detail in Fig. 5. \textbf{(c)} PINEM image of the PIP generated by a linear array of short NCs (measured lengths $\simeq$ 0.6, 0.7, 0.8 and \SI{0.9}{\micro\meter}, respectively, widths $\simeq$ 275 nm), photoexcited by linearly polarized light at $\varphi=-78^{\circ}$. \textbf{(d)} PINEM image of the PIP generated by two vertically offset linear arrays of NCs (length $\simeq$ \SI{2.1}{\micro\meter}, width $\simeq$ 270 nm), photoexcited by linearly polarized light at $\varphi=-20^{\circ}$. The resulting SPP standing wave interference shows a periodicity of $\lambda_{\text{SPP}}$/2 along the vertical (counter-)propagation direction (dashed horizontal lines), yielding an estimated SPP wavelength of $\lambda_{\text{SPP, exp}}$ = 638 $\pm$ 32 nm. The dashed elliptical shape depicts the approximate outline of the electron beam footprint. All panels are recorded at $\tau =$ 0 using $\lambda_{\text{0}}$ = 786 nm and share the linear false color scale depicted in the middle.}
\end{figure}
\end{center}
\twocolumngrid
\section{Discussion and Outlook}
\indent The unique combination of characterization and control capabilities offered by PINEM enables new scenarios for nanoplasmonic circuits. Advanced nanopatterning and multilayered designs could be used to exploit the quantum entanglement of plasmons \cite{altewischer02,fakonas14}, efficient plasmon sources based on the quantum \v{C}erenkov effect\cite{kaminer15,genevet15} or local nonlinear effects at topological defects\cite{yin14}, to name a few. The emergence of 2D solids such as graphene\cite{novoselov12}, hexagonal BN\cite{pakdel14} and the layered transition metal dichalcogenides\cite{wang12b}, their integration into heterostructures\cite{britnell13,gong14}, and the availability of new plasmonic materials\cite{wang15} offer a comprehensive set of tools to develop and advance plasmon-based technology in a wide range of areas, including plasmonic sensing\cite{stewart08}, nonlinear plasmonics\cite{kauranen12}, and plasmonic lasers (SPACERS)\cite{zhou13}. Furthermore, the ability to control the spatial shape and temporal evolution of confined plasmonic fields enables the implementation of nanoscale experiments to showcase fundamental properties of quantum mechanics\cite{piazza15} and coherently manipulate electron beams\cite{feist15}. In analogy to the Kapitza-Dirac effect\cite{kapitza33,freimund01}, describing the interaction between photons and electrons in free space, it was recently shown in Ref. \citenum{garcia16} that electron diffraction from standing plasmon waves can be feasible under experimental conditions similar to those achieved in PINEM, paving the way to the active control of electron diffraction for the generation of tailored beams (e.g. vortex beams) and other exotic superpositions of electron wave-functions.\newline
%
\indent Summarizing, in this work we demonstrate the \emph{in-situ} visualization of photoinduced plasmonic interference patterns confined to otherwise inaccessible buried interfaces, enabling a critical tool for the investigation and development of complex plasmonic heterostructures and advanced multilayer devices. Our experiments demonstrate the feasibility of ultrafast imaging of plasmon dynamics using PINEM, allowing us to experimentally measure the carrier wavelength and propagation speed of SPPs traveling at buried interfaces directly in the time domain. Furthermore, we show that transient plasmonic interference patterns can be shaped, manipulated and controlled through both the polarization of the excitation light and the nanopatterning architecture, thus facilitating a widely tunable range of nanoscale near-field structures. Finally, the results presented here represent a considerable advance toward the realization of the recently proposed methodology involving inelastic electron diffraction from transient plasmonic gratings.
\section*{Methods}

\subsection{Materials.}
A silver thin film was sputtered onto an in-house fabricated \ce{Si3N4}-on-Si support using an EMITECH K575x sputter coater equipped with an Ag target (100 mA, 60s exposure, 3 $\times$ 10$^{-4}$ mbar base pressure).  The nominal thickness of the \ce{Si3N4} membrane was 50 nm, with potential thickness variation between 35 and 50 nm. Nanopatterns were written in the Ag layer using a raster-scanned focused-ion beam (FEI Nova Nanolab 200 FIB/SEM) with typical beam currents of 9-10 pA at a 30 kV voltage. Supplementary Figure 1b-c shows an SEM characterization of the sample, which was kept in low oxygen conditions and measured in the UTEM (at 295 K and $\sim$ 10$^{-5}$ Pa) within a few days of patterning in the FIB. For the propagation experiment, nominal NC widths were 250 nm (experimentally measured (SEM) widths $\simeq$ 274 nm) with nominal NC lengths incrementing from 700 to 1300 nm in 100 nm steps (from left to right, NCs in field of view in Fig. 3a, experimentally measured (SEM) lengths ($\pm 0.3 \%$): 698, 809, 919, 982, 1113, 1175, and 1273 nm).

\subsection{Experimental.}
A 300 kHz train of linearly polarized light pulses was split to yield two beams, one of which was frequency-tripled to deliver few-nJ UV pulses (262 nm) to a custom truncated-cone \ce{LaB6} photo-cathode (\SI{15}{\micro\meter} diameter flat tip, produced by AP-Tech) in a custom-modified ultrafast transmission electron microscope (UTEM)\cite{piazza13}. The UV pulse train was attenuated to \SI{}{\micro\watt}-range average power to ensure photoemission of ultrafast 200 keV single-electron bunches, minimizing electron-electron interactions in the probe and yielding an optimized beam coherence at an electron bunch duration of $\sim$ 650 fs FWHM (bunch envelope proportional to $e^{-(\frac{t+\tau}{\Delta_{e}})^2}$, where $\tau$ is the relative time delay of the electron bunch with respect to the optical pump pulse, see Figure 1). The second optical beam (FWHM duration $\sim$ 105 fs, central wavelength
$\lambda_{\text{0}}$ = 786 $\pm$ 15 nm), after undergoing a variable delay, was weakly focused on the sample in the UTEM at near-normal incidence using an achromatic doublet ($f$ = 250 mm) to a spot size of $\simeq$ \SI{100}{\micro\meter} diameter at the sample plane, such that the electric field of the pump beam was uniform across the field of view of the electron beam. The time delay $\tau$ is defined as the arrival time of the midway point of the Gaussian electron probe pulse at the sample $z$-position with respect to that of the Gaussian optical pump pulse. Optical pump fluences employed were of the order of 1 to 4 mJ cm$^{-2}$, corresponding to a peak excitation energy density of $\simeq$ 50 GW cm$^{-2}$. Under these experimental conditions, the generation of SPPs by the transient electric field associated with the electron probe beam is a much weaker effect, which can be considered negligible\cite{park10}. A detailed description and characterization of the modified JEOL JEM 2100 microscope can be found elsewhere\cite{piazza13}. For the PINEM experiments described in this work, the UTEM was operated at 200 keV in photoelectron mode. The GIF imaging camera was operated with a 0.05 eV-per-channel dispersion setting, while typical exposure times of the 2048$\times$2048 pixel CCD sensor were 60 s for images and 10 s for spectra. Electron energy loss spectra were aligned based on their zero loss peak (ZLP) positions using a differential-based maximum intensity alignment algorithm.

\subsection{Image processing.}
Variable delay PINEM frames were spatially aligned using a custom video stabilization algorithm. All obtained PINEM images were median-filtered for single pixel noise and spike removal, and background subtracted using a high-pass spatial-frequency Fourier filter. PINEM image calibration was done using the SEM-measured NC lengths in the field of view as independent references, with the final nm-per-pixel parameter taken as the mean of the calibration factors independently determined for the different NCs. The resulting uncertainty in the image scale is the principal contribution to the error bars on the extracted $\lambda_{\text{SPP, exp}}$ and $k_{\text{SPP, exp}}$ values. The fringe positions in Figures 4d and 5a were determined by fitting Gaussian lineshapes to the corresponding integrated spatial traces (the image in Fig. 4d was horizontally integrated), and the resulting $\lambda_{\text{SPP, exp}}$ values were calculated from the average fringe spacing. Light polarization angles $\varphi$ at the sample position inside the UTEM were determined to be within a systematic error of $\pm$ 5$^{\circ}$ and verified by comparison to simulations of the plasmonic field.\newline
\indent For a robust determination of the SPP group velocity, the positive $y$-range of PINEM images recorded at different time delays $\tau$ were divided up into a series of discrete horizontal slices, according to various slicing schemes. The data in Fig. 3b correspond to the spatial FFT analysis of a series of slices with a $y$-range of $\Delta y$ = \SI{0.42}{\micro\metre}. For each slice, a Gaussian temporal profile was fitted to the $\tau$-dependence of the integrated intensity of the spatial FFT features characteristic of the SPP interference. Full slicing analyses were carried out for slicing schemes having a $\Delta y$ = 0.14, 0.28, 0.42, 0.56 and \SI{0.70}{\micro\metre}, in each case with a $y$-center offset equal to 0.5$\Delta y$ (50\% slice overlap). The resulting data points (fitted temporal Gaussian center vs. $y$-range center) for each of the slices in the different slicing schemes were combined and fitted with a single straight line, whose slope represents the group velocity of the propagating plasmonic wave (see Fig. 3c).

\subsection{Analytical calculations.}
The dispersion diagram in Fig. 2d represents the energy-loss- and in-plane-momentum-transfer-resolved loss probability experienced by an electron crossing the Ag/\ce{Si3N4} structure of the experimental sample under normal incidence. The probability is calculated analytically following the methods described in Ref. \citenum{garcia04}. In brief, (1) the total energy loss is obtained by integrating along the trajectory of the retarding force produced by the induced electric field that is created by the electron and acting back on it; (2) the result is expressed as a double integral over energy- ($\Delta E$) and momentum-transfer ($k_{\parallel}$) components, which are interpreted as $\Delta E$ times the energy-and-momentum-resolved distribution of the loss probability (i.e., the quantity actually represented in the figure); (3) the induced field is in turn obtained by considering a \emph{source} field produced by the electron in each of the media as if it were moving in an extended homogeneous material; (4) this source field is then scattered at each of the interfaces, where it produces TM reflected waves (note that the source field does not couple to TE waves due to the $m$=0 axial symmetry of the problem for normal electron incidence); (5) the amplitudes of these reflected waves are determined by the continuity of the electric field and the normal displacement; (6) finally, the integral of the retarding force along the trajectory is carried out analytically within each material for every combination of $\Delta E$ and $k_{\parallel}$. The materials (Ag and \ce{Si3N4}) are represented through their dielectric functions, taken from tabulated measured data (Refs. \citenum{johnson72} and \citenum{palik98}, respectively).\newline 
\indent An approximate plasmonic interference pattern for the nanostructure imaged in Figure 4d was analytically calculated based on arrays of simple point dipole SPP sources (see Supplementary Figure 3). The model takes into
account the finite propagation distance $\gamma$ $\approx$ \SI{64}{micron}, defined as the $1/e$ decay distance of the plasmon electric field, calculated from the tabulated permittivities for SPPs at a Ag/\ce{Si3N4} interface (semi-infinite slabs) at the photon energy under consideration. This value is consistent with the experimental observations in this work, which show little spatial plasmonic field decay within the field of view. All sources were assumed to be monochromatic ($\lambda_{\text{SPP}}$ = 638 nm), radiating SPP waves travelling radially outward in a dipolar pattern. All dipoles were oriented along the same direction (maximum radiated field along $\varphi=-20^{\circ}$, with the point dipole oscillating along the perpendicular direction, i.e. $\varphi=70^{\circ}$). The resulting complex $E_{\text{z}}$-field components originating from the different sources were coherently summed, after which uniform time-averaging was incorporated by taking the square modulus of the sum.
%


\clearpage

\onecolumngrid
\section*{Supplementary Figures}
\def\figurename{Supplementary Figure}
\setcounter{figure}{0}
\begin{center}
\begin{figure}[ht]
\includegraphics[width=14cm]{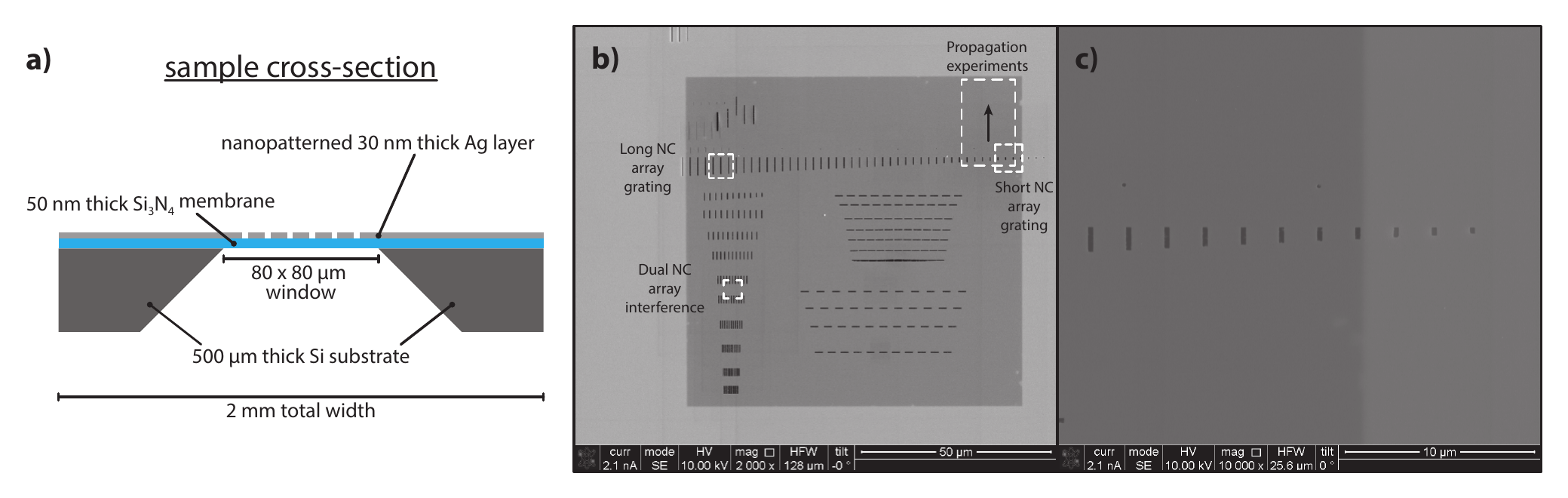}
\caption{\label{SuppFig1} The nanopatterned Ag-on-Si3N4 sample. (a) Schematic cross-section of the sample. (b) Overview SEM image of the various perforating nanocavities written in the Ag film. Areas imaged using the PINEM technique in this work are indicated by labeled dashed rectangles. (c) SEM Zoom-in on the NC array imaged in the propagation experiment described in figures 2 and 3 of the main text.}
\end{figure}
\end{center}
\begin{center}
\begin{figure}[htb]
\includegraphics[width=14cm]{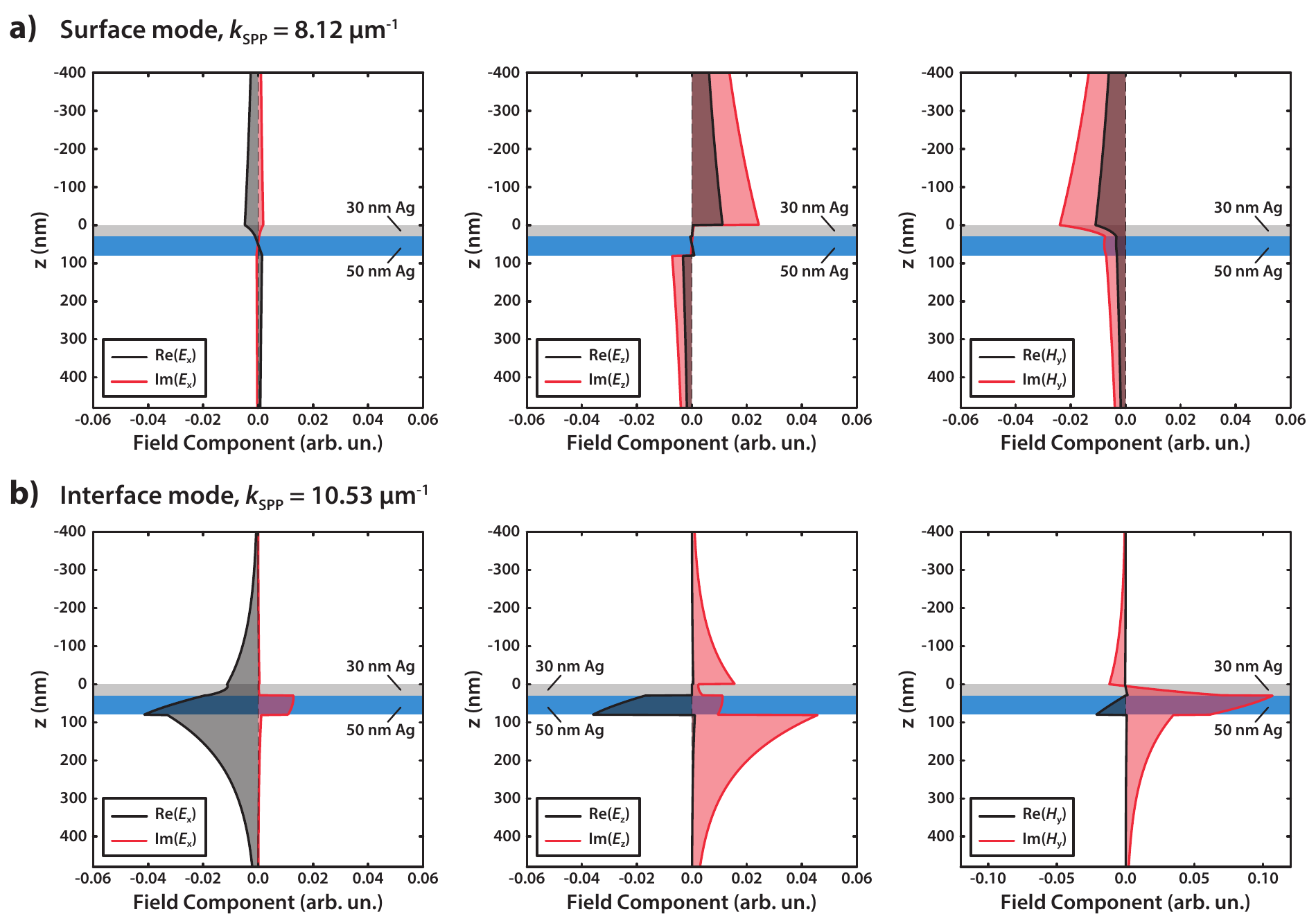}
\caption{\label{SuppFig2} Analytically calculated mode profiles for the two distinct plasmon modes identified in Figure 2d. The $z$-dependence ($z$ corresponds to the direction perpendicular to the sample surface) of the relevant electromagnetic components for the surface (panel (a)) and buried interface (panel (b)) SPP modes, excited at an energy of 1.577 eV, are plotted in relative arbitrary units.}
\end{figure}
\end{center}
\begin{center}
\begin{figure}[htb]
\includegraphics[width=9cm]{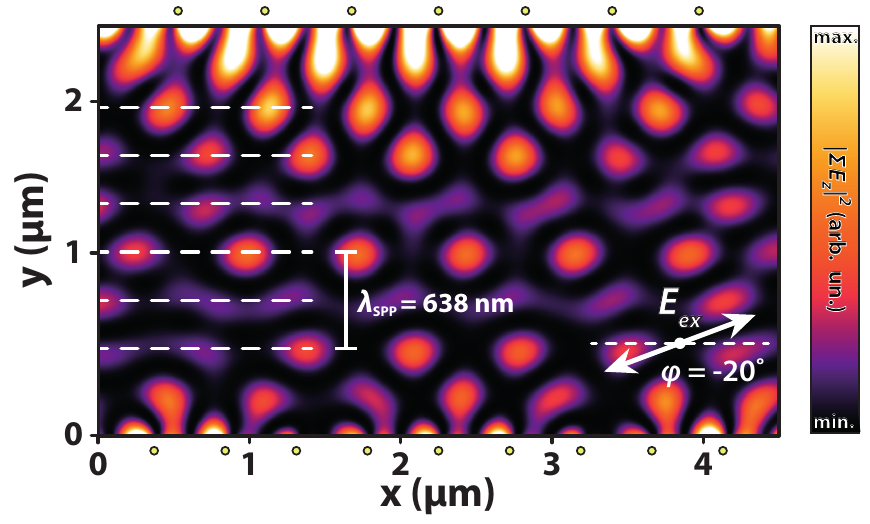}
\caption{\label{SuppFig3} Analytically calculated plasmonic interference pattern based on simple point dipole SPP sources (relative positions indicated by filled circles at panel top and bottom). Dipoles in each array are of equal strength, with their maximum radiated field oriented along the double-headed white arrow (i.e. the corresponding dipole oscillates along the direction perpendicular to the arrow). Plotted is the squared modulus of the linear superposition of $E_{\text{z}}$-components radiated by the different sources. Calculated with corresponding geometry and a carrier wavelength of $\lambda_{\text{SPP}}$ = 638 nm, this simple dipole model shows excellent agreement with the experimental PINEM image of figure 4D, including the alternating pattern of continuous intensity wiggles, the arrays of intensity islands and the $\lambda_{\text{SPP}}$/2 periodicity along the $y$ direction.}
\end{figure}
\end{center}
\end{document}